\RequirePackage{silence}
\documentclass[fleqn,usenatbib,useAMS]{mnras} 
\usepackage{graphicx}

\let\normalbigoplus \bigoplus
\usepackage{amsmath}
\let\bigoplus \normalbigoplus

\usepackage{amsfonts}
\usepackage{float}
\usepackage{bm}
\usepackage[perpage,symbol*]{footmisc}
\usepackage{ae,aecompl}
\usepackage{array}
\usepackage{soul}
\usepackage{mathtools}
\usepackage{multirow}

\newcommand{\appropto}{\mathrel{\vcenter{
		\offinterlineskip\halign{\hfil$##$\cr 
			\propto\cr\noalign{\kern2pt}\sim\cr\noalign{\kern-2pt}}}}}

\title{Directly testing gravity with Proxima Centauri} 

\author[Indranil Banik \& Pavel Kroupa]{Indranil Banik$^{1}$\thanks{Email:
\href{mailto:ibanik@astro.uni-bonn.de}{ibanik@astro.uni-bonn.de} (Indranil Banik)\newline $~~~~~~~~~~~~~~~~~$ \href{mailto:pavel@astro.uni-bonn.de}{pavel@astro.uni-bonn.de} (Pavel Kroupa)} and Pavel Kroupa$^{1,2}$\\
$^{1}$Helmholtz-Institut f\"ur Strahlen und Kernphysik (HISKP), University of Bonn, Nussallee 14$-$16, D-53115 Bonn, Germany \\
$^{2}$Charles University, Faculty of Mathematics and Physics, Astronomical Institute, V Hole\v{s}ovi\v{c}k\'ach 2, CZ-18000 Praha 8, Czech Republic}

\pubyear{2019}
\begin{document}
\label{firstpage}
\pagerange{\pageref{firstpage}--\pageref{lastpage}}

\maketitle

\begin{abstract}

The wide binary orbit of Proxima Centauri around $\alpha$ Centauri A and B differs significantly between Newtonian and Milgromian dynamics (MOND). By combining previous calculations of this effect with mock observations generated using a Monte Carlo procedure, we show that this prediction can be tested using high precision astrometry of Proxima Centauri. This requires ${\approx 10}$ years of observations at an individual epoch precision of $0.5 \, \mu$as, within the design specifications of the proposed Theia mission. In general, the required duration should scale as the 2/5 power of the astrometric precision.

A long-period planet could produce a MOND-like astrometric signal, but only if it has a particular ratio of mass to separation squared and a sky position close to the line segment connecting Proxima Centauri with $\alpha$ Centauri. Uncertainties in perspective effects should be small enough for this test if the absolute radial velocity of Proxima Centauri can be measured to within ${\approx 10}$ m/s, better than the present accuracy of 32 m/s. We expect the required improvement to become feasible using radial velocity zero points estimated from larger samples of close binaries, with the Sun providing an anchor.

We demonstrate that possible astrometric microlensing of Proxima Centauri is unlikely to affect the results. We also discuss why it should be possible to find sufficiently astrometrically stable reference stars. Adequately addressing these and other issues would enable a decisive test of gravity in the currently little explored low acceleration regime relevant to the dynamical discrepancies in galactic outskirts.


\end{abstract}


\section{Introduction}
\label{Introduction}

One of the great mysteries in contemporary astronomy concerns the true cause of the very large dynamical discrepancies between the observed rotation curves of galaxies and the predictions of Newtonian gravity applied to their luminous matter distributions \citep[e.g.][]{Babcock_1939, Rubin_Ford_1970, Rogstad_1972}. These acceleration discrepancies are usually attributed to halos of cold dark matter surrounding each galaxy \citep{Ostriker_Peebles_1973}. However, the discrepancies follow some remarkable regularities \citep{Famaey_McGaugh_2012} that can be summarised as a unique relation between the acceleration inferred from the rotation curve and that expected from the baryonic distribution \citep{McGaugh_Lelli_2016}. Such a radial acceleration relation (RAR) was predicted several decades earlier using Milgromian dynamics \citep[MOND,][]{Milgrom_1983}. In this model, the dynamical effects usually attributed to dark matter are instead provided by an acceleration dependence of the gravity law. The gravitational field strength $g$ at distance $r$ from an isolated point mass $M$ transitions from the Newtonian ${GM/r^2}$ law at short range to
\begin{eqnarray}
g ~=~ \frac{\sqrt{GMa_{_0}}}{r} ~~~\text{for } ~ r \gg \overbrace{\sqrt{\frac{GM}{a_{_0}}}}^{r_{_M}} \, .
\label{Deep_MOND_limit}
\end{eqnarray}
MOND introduces $a_{_0}$ as a fundamental acceleration scale of nature below which the deviation from Newtonian dynamics becomes significant. Empirically, $a_{_0} \approx 1.2 \times {10}^{-10}$ m/s$^2$ to match galaxy rotation curves \citep{Begeman_1991, McGaugh_2011}. Remarkably, this is the same order of magnitude as the acceleration at which the classical energy density of a gravitational field \citep[][equation 9]{Peters_1981} becomes comparable to the dark energy density $u_{_\Lambda} \equiv \rho_{_\Lambda} c^2$ that conventionally explains the accelerating expansion of the Universe \citep{Ostriker_Steinhardt_1995, Riess_1998, Perlmutter_1999}.
\begin{eqnarray}
\frac{g^2}{8\mathrm{\pi}G} ~<~ u_{_\Lambda} ~~\Leftrightarrow~~ g ~\la~ 2\mathrm{\pi}a_{_0} \, .
\label{MOND_quantum_link}
\end{eqnarray}
MOND could thus be a result of poorly understood quantum gravity effects \citep[e.g.][]{Milgrom_1999, Pazy_2013, Verlinde_2016, Smolin_2017, Bagchi_2019}.

Regardless of its underlying microphysical explanation, MOND can accurately match the rotation curves of a wide variety of both spiral and elliptical galaxies across a vast range in mass, surface brightness and gas fraction using only the distribution of luminous matter \citep[][figure 5]{Lelli_2017}. Fits to individual rotation curves show that intrinsic scatter about its predictions must be ${<13\%}$ and is consistent with 0 \citep{Li_2018}. A few discrepant galaxies were claimed by \citet{Rodrigues_2018}, but it was later shown that the actual discrepancies are either very mild or arise when the distance is particularly uncertain and could plausibly be outside the range they allow \citep{Kroupa_2018}. Given that most of this data was obtained after the MOND field equation was first published \citep{Bekenstein_Milgrom_1984}, these achievements are successful a priori predictions. It is difficult to explain the success of these predictions in a conventional gravity context, even with the observational facts in hand \citep{Desmond_2016, Desmond_2017}.

In principle, a galaxy is large enough that it is physically possible to fit in the amount of dark matter required to explain its observed rotation curve, even if the required distribution was not predicted. However, this approach causes tension with observational data, in particular the apparent absence of dynamical friction expected to arise between the postulated massive and extended dark matter halos \citep{Angus_2011, Kroupa_2012, Kroupa_2015, Oehm_2017}. Another problem is that several Local Group dwarf galaxies have anomalously high radial velocities compared to a detailed three-dimensional timing argument calculation \citep{Peebles_2017, Banik_2017_anisotropy}. The latter authors showed that these dwarfs are more naturally explained in MOND due to its implication that there was a past close flyby between the Milky Way and Andromeda galaxies ${\approx 8}$ Gyr ago \citep{Zhao_2013, Bilek_2018, Banik_Ryan_2018}. Three-body interactions with a nearby dwarf galaxy could fling it outwards at high speed.

Because MOND is an acceleration-dependent theory, its effects could become apparent in a rather small system if this has a sufficiently low mass (Equation \ref{Deep_MOND_limit}). In fact, the MOND radius $r_{_M}$ is only 7000 astronomical units (7 kAU) for a system with $M = M_\odot$. This suggests that the orbits of distant Solar System objects might be affected by MOND \citep{Pauco_2016}, possibly accounting for certain correlations in their properties \citep{Pauco_2017}. However, it is difficult to accurately constrain the dynamics of objects at such large distances.

Such constraints could be obtained more easily around other stars if they have distant binary companions. As first suggested by \citet{Hernandez_2012}, the orbital motions of these wide binaries (WBs) should be faster in MOND than in Newtonian gravity. Moreover, it is likely that many such systems would form \citep{Kouwenhoven_2010, Tokovinin_2017}. Indeed, data from the Gaia mission \citep{Perryman_2001} strongly suggests the presence of several thousand WBs within $\approx 150$ pc \citep{Andrews_2017}. The candidate systems they identified are mostly genuine, with a contamination rate of ${\approx 6\%}$ \citep{Andrews_2018} estimated using the second data release of the Gaia mission \citep[Gaia DR2,][]{Gaia_2018}.

The wide binary test (WBT) of gravity was considered in more detail by \citet{Pittordis_2018}, who set up simulations of WBs in Newtonian gravity and several theories of modified gravity, including MOND. These calculations were revisited by \citet{Banik_2018_Centauri} using self-consistent MOND simulations that include the external field from the rest of our Galaxy and use an interpolating function consistent with the RAR (see their sections 2.1 and 7.1, respectively). Their main result was that MOND enhances the orbital velocities of Solar neighbourhood WBs by ${\approx 20\%}$ above Newtonian expectations, consistent with their analytic estimate \citep[see their section 2.2 and][]{Banik_2015}. Using statistical methods they developed, they showed that $\approx 500$ WB systems would be required to detect this effect if measurement errors are neglected but only sky-projected quantities are used (these are expected to be more accurate).

The WBT was first attempted by \citet{Hernandez_2012} using the WB catalogue of \citet{Shaya_2011}, who analysed Hipparcos data with Bayesian methods to identify WBs within 100 pc \citep{Leeuwen_2007}. Typical relative velocities between WB stars seemed to remain constant with increasing separation instead of following the expected Keplerian decline \citep[][figure 1]{Hernandez_2012}. Recently, this work was revisited by \citet{Hernandez_2018} using data from Gaia DR2. However, some problems with the analysis were soon pointed out \citep{Badry_2019}. A more careful analysis found no clear evidence of a departure from Newtonian expectations beyond the MOND radius \citep{Banik_2019}. This analysis is still very preliminary and certainly does not rule out the expected MOND effect. But it does show that Gaia DR2 results are likely accurate enough to enable the WBT if this is done using only one component of $\bm{v}_{rel}$ \citep[][section 3.2]{Shaya_2011}. The use of less data would double the required number of WBs to ${\approx 1000}$, but this is still much less than the number of systems in the \citet{Andrews_2018} catalogue. Conducting the WBT this way minimises the impact of distance uncertainties.

The main feature of the WBT is that the law of gravity is constrained using relative velocities between stars. Ideally, the acceleration itself would be directly measured. The typical acceleration scale of this problem is $a_{_0}$, which implies a deviation from uniform rectilinear motion of $\frac{1}{2} a_{_0} t^2 = 6000$ km over $t = 10$ years. This rather small length is currently not possible to resolve at interstellar distances. However, rapid technological progress motivates us to consider the prospects for detecting WB orbital accelerations in the future using Proxima Centauri (P Cen), a component of our nearest WB and also the nearest star to the Sun. P Cen is one of the most active M dwarfs \citep{Kroupa_1989} and orbits the close (24 AU) binary $\alpha$ Cen A and B at a distance of 13 kAU \citep{Kervella_2017}. The P Cen orbit would thus be significantly affected by MOND \citep{Beech_2009, Beech_2011}. Therefore, it is likely to be the first WB whose relative acceleration will be determined \citep[][section 5.2.3]{Kervella_2018}. Though unlikely in the Gaia era, we consider whether this challenging measurement can be taken by the proposed Theia mission \citep{Theia_2017}.

We begin with a careful calculation of P Cen's expected anomalous acceleration in MOND (Section \ref{Expected_acceleration}) and the amount of time required to detect this at different levels of astrometric precision (Section \ref{Detectability}). We then consider various systematics which may limit our ability to decisively test the MOND prediction (Section \ref{Systematics}). Our conclusions are presented in Section \ref{Conclusions}.

\section{Detecting the anomalous acceleration of Proxima Centauri}

\subsection{The expected acceleration}
\label{Expected_acceleration}

The orbital acceleration of P Cen is expected to differ significantly between Newtonian and Milgromian dynamics. The detailed calculations were presented in \citet[][section 9.1]{Banik_2018_Centauri}. There, it was assumed that P Cen is a test particle orbiting $\alpha$ Cen A and B, which was treated as a single point mass $\alpha$ Cen. This is reasonable given the very small separation of the $\alpha$ Cen close binary and the fact that its mass is ${\approx 17 \times}$ larger than that of P Cen \citep[][table 1]{Kervella_2017}.

In principle, our analysis should consider the fact that P Cen constitutes a non-zero fraction of the total system mass. However, figure 8 of \citet{Banik_2018_Centauri} showed that this is expected to affect the $\alpha$ Cen-P Cen relative acceleration by only a few percent even if P Cen constituted 30 percent of the total system mass. Their section 7.3 explains why this is the case $-$ in the external field dominated regime, forces can be superposed just like in Newtonian gravity, so the mass ratio does not affect the relative acceleration. In general, the mutual acceleration of a two-body system is slightly reduced in MOND if the mass is split more equally between its components \citep{Zhao_2010}. We approximately account for this by not including the mass of P Cen in the total mass of the system.

Theoretical uncertainties of a few percent are inevitable because the MOND interpolating function is not perfectly known. But observational improvements may ultimately necessitate more precise calculations on a three-dimensional grid, which could be done using the publicly available MOND grid solver Phantom of RAMSES \citep[PoR,][]{PoR}. The interpolating function is likely to be the main uncertainty going forwards since the mass of $\alpha$ Cen is known to much better than 1 percent \citep[][table 1]{Kervella_2016} and even the P Cen mass is known to within 2 percent \citep[][table 1]{Kervella_2017}.

Unlike in Newtonian gravity, the Milgromian acceleration of P Cen is not directly towards $\alpha$ Cen but is instead directed $3.1^\circ$ away due to the non-linearity of MOND and the external Galactic gravitational field on the system \citep[a non-radial gravity law is demonstrated analytically in][]{Banik_2015}. The main effect of MOND is that the expected acceleration is 45\% larger than in Newtonian dynamics ($0.87 \, a_{_0}$ instead of $0.60 \, a_{_0}$).

This extra acceleration would have a small effect on observables if perspective effects are accurately accounted for (Section \ref{Perspective_effect}) and there is no planet with the exact properties required to mimic the MOND signal (Section \ref{Undetected_companions}). After a decade of observations, the radial velocity, right ascension and declination would all exceed Newtonian expectations by 0.52 cm/s, 6.60 $\mu$as and 2.83 $\mu$as, respectively, implying a total astrometric anomaly of 7.18 $\mu$as. In the rest of this contribution, we neglect the effect on the radial velocity since it is very small. This is because the radial velocity only changes linearly with time in the presence of some anomalous acceleration, whereas the sky position would change quadratically.

\subsection{Detectability of the effect}
\label{Detectability}

Although the expected effect is very small, it grows rather quickly with time $T$. Assuming observations are carried out regularly with equal accuracy, the number of observations grows with $T$ so the `uncertainty' drops as $T^{-1/2}$. Because the signal itself $\propto T^2$, the signal to noise ratio grows as $T^{5/2}$. This is much faster than the $T^{3/2}$ improvement expected for the measurement of a constant proper motion because this implies only a linear change in sky position with time.

We now consider in more detail the duration $T$ required to detect a significant deviation from the Newtonian model given an individual astrometric pointing accuracy $\sigma$. For this purpose, we define orthogonal directions $x$ and $y$ on the sky plane at the position of P Cen. $y$ points in the sky-projected direction of the extra acceleration predicted by MOND compared to Newtonian dynamics.

Using this co-ordinate system, we perform $10^6$  Monte Carlo trials in which the observed astrometric (MOND $-$ Newton) residuals of $x$ and $y$ at times $t_i$ are found under the assumption that all known effects have been subtracted from the sky positions. These include the initial position, proper motion and parallax of P Cen, the perspective effect arising from its radial and tangential velocity (Section \ref{Perspective_effect}), the Galactic acceleration of the whole system relative to that of the Sun (Section \ref{Galactic_acceleration}) and the expected orbital acceleration of P Cen in Newtonian dynamics. After subtracting these effects, we expect that
\begin{eqnarray}
	x_i ~&=&~ \delta \, ,\\
	y_i ~&=&~ 7.18 \, \mu\text{as} \left( \frac{t_i}{10 ~\text{yr}}\right)^2 ~+~ \delta \, ,
\end{eqnarray}
where $\delta$ is a Gaussian random variable with dispersion $\sigma/\sqrt{2}$. The value of $\delta$ is assumed to be independent for $x$ and $y$, though in reality some correlation might be expected. For Gaia data, this mainly arises due to different astrometric uncertainties in the along-scan and across-scan directions. However, Theia should directly observe positions on the sky plane, making this issue much less significant.

We create a list of $\left(t_i, x_i, y_i \right)$ based on the observing epochs $t_i$ having equal spacing and covering a total duration of 50 years. To see how feasible this test is after a shorter period of e.g. 10 years, we simply restrict the remainder of our analysis (see below) to the first 10 years of observations.

To some extent, any anomalous acceleration is degenerate with changes to the initial position and velocity of P Cen. Thus, we fit a line to $\left(t_i, x_i \right)$ of the form $x_i = at_i + b$, where the fitting parameters $a$ and $b$ are found in the standard way.
\begin{eqnarray}
	\widetilde{t}_i \, &\equiv& \, t_i - \overline{t} \, ,~~\widetilde{x}_i \, \equiv \, x_i - \overline{x} \, , \nonumber \\
	a ~&=&~ \frac{\sum_{i = 1}^N{\widetilde{t}_i \widetilde{x}_i}}{\sum_{i = 1}^N{\widetilde{t}_i \widetilde{t}_i}} \, , \nonumber \\
	b ~&=&~ \overline{x} - a\overline{t} \, .
	\label{Linear_fit}
\end{eqnarray}
Here, we use $\overline{q}$ to denote the mean value of any quantity $q$ over the $N$ measurements that we consider. Once the mean values of $t$ and $x$ are subtracted, it is easy to find the best-fitting slope $a$ and intercept $b$ required for the best-fitting line to pass through the mean of the mock measurements. This is valid regardless of the measurement uncertainty, as long as it is the same at all epochs. We use a similar procedure to find the best-fitting line to $\left(t_i, y_i \right)$ since there is no a priori reason to expect any correlation between the $x_i$ and $y_i$.

Once the best-fitting lines have been obtained, we determine their $\chi^2$ with respect to the observations for a measurement uncertainty of $\sigma/\sqrt{2}$ along each direction. For $N$ observing epochs, we have ${\left(2N - 4\right)}$ degrees of freedom because 4 quantities are required to specify the initial astrometric position and proper motion. Without the MOND term, we thus expect that $\chi^2 = {\left(2N - 4\right)}$ on average. Using the $chi2cdf$ function in \textsc{matlab}$^\text{\textregistered}$, we determine the maximum value of $\chi^2$ for $\left(2N - 4\right)$ degrees of freedom at some confidence level. We take this to be $5.73 \times 10^{-7}$, the two-tailed probability outside the central 5 sigma of a Gaussian distribution. To reliably detect the signature of MOND, it is necessary that a significant fraction $P_{detection}$ of our Monte Carlo trials yield a $\chi^2$ exceeding this value. $P_{detection}$ can be thought of as the probability of detecting a significant departure from Newtonian expectations.

\subsubsection{Analytic estimate of the required precision}
\label{Analytic_estimate}

Before showing the results of our numerical calculations, it is worth considering the problem analytically. The problem can be phrased as finding the $\chi^2$ of the best linear fit to the parabola $y = t^2$ over the range ${t = 0 - 1}$. The best fit line can be found using Equation \ref{Linear_fit}.
\begin{eqnarray}
	y ~=~ t \, - \, \frac{1}{6} \, .
	\label{Linear_fit_parabola}
\end{eqnarray}

The goodness of this fit can be quantified by integrating the squared error.
\begin{eqnarray}
	\int_0^1 \left[ t^2 - \left(t - \frac{1}{6} \right) \right]^2 dt ~=~ \frac{1}{180} \, .
	\label{Error_linear_fit}
\end{eqnarray}

In general, measurements deviate from Equation \ref{Linear_fit_parabola} both because it is the incorrect model and because there are measurement uncertainties. For any constant $a$ and random variable $X$ with mean 0, the expected value of $\left( X + a \right)^2$ exceeds the variance of $X$ by $a^2$. Therefore, we expect that the total $\chi^2$ of the model is on average $2N + \frac{N}{180}$ after $N$ observing epochs if the astrometric anomaly is equal to the one-dimensional measurement uncertainty $\sigma/\sqrt{2}$. This is valid if $N \gg 4$ so that we have much more data points than the 4 model parameters.

Assuming that observations are carried out over 10 years with 3 epochs per year so that ${N = 30}$, we get that on average $\chi^2$ exceeds the expected value by
\begin{eqnarray}
	\Delta \chi^2 ~\approx~ \frac{30}{180} \left( \frac{7.18 \, \mu as}{\sigma/\sqrt{2}} \right)^2
	\label{Delta_chi_sq}
\end{eqnarray}

The total $\chi^2$ is expected to be 60, so a significant detection of MOND effects requires that $\Delta \chi^2 \gg 60$. In particular, for 60 degrees of freedom, the $\chi^2$ distribution can be fairly well approximated by a Gaussian with mean 60 and dispersion $\sqrt{120}$. To achieve a 5 sigma departure from Newtonian dynamics most of the time, we require that $\Delta \chi^2 = 5\sqrt{120}$. But to ensure that a null detection provides strong evidence against MOND, it must be very likely to observe a 5 sigma departure from Newtonian expectations if Milgromian dynamics were correct. Thus, $5 \sqrt{120}$ must be significantly smaller than the $\Delta \chi^2$ estimated in Equation \ref{Delta_chi_sq}. This is possible by using a stricter condition e.g.
\begin{eqnarray}
	\Delta \chi^2 ~\geq~ 10\sqrt{60} \, .
	\label{Required_delta_chi_sq}
\end{eqnarray}
This approximately corresponds to a null detection of MOND ruling it out at $5\sigma$.

Equations \ref{Delta_chi_sq} and \ref{Required_delta_chi_sq} can be combined to show that
\begin{eqnarray}
	\sigma ~\le~ 0.33 \, \mu as \, .
\end{eqnarray}
We therefore estimate that an astrometric precision of $\approx 0.5 \, \mu$as is required to reliably distinguish Newtonian from Milgromian gravitation after 10 years using ${f = 3}$ observations per year. Our discussion at the start of Section \ref{Detectability} indicates that the signal to noise ratio should be $\propto T^{5/2}f^{1/2}$. Thus, a higher observing frequency $f$ reduces the required duration $T$ according to $T \propto f^{-1/5}$. Consequently, our results should not be very sensitive to the exact details of the observing strategy but do depend strongly on the total observing duration $T$.

\subsubsection{Results of numerical simulations}
\label{Numerical_results}

\begin{figure}
	\centering
	\includegraphics[width = 8.5cm] {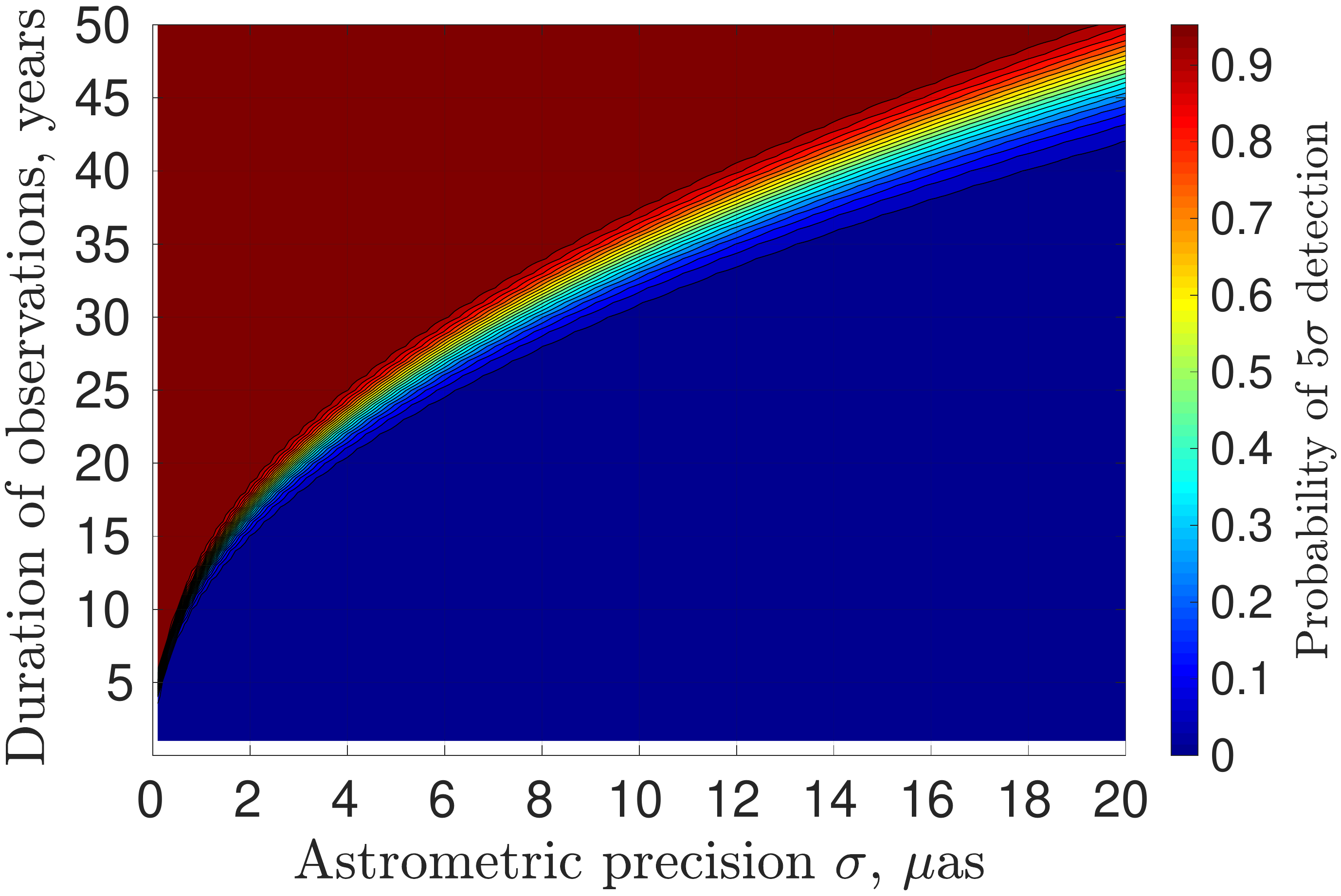}
	\caption{The proportion of Monte Carlo trials yielding a 5 sigma discrepancy with Newtonian model predictions if the true dynamics are Milgromian and 3 observations are taken each year. For a given astrometric precision, the detection probability rises rather sharply from ${\approx 0}$ to ${\approx 1}$ after a certain amount of time, which can be considered the required duration for the experiment to yield conclusive results. This duration is shown in Figure \ref{Required_duration_5_sigma} for a threshold of 99\%.}
	\label{P_detection_3}
\end{figure}

\begin{figure*}
	\centering
	\includegraphics[width = 8.5cm] {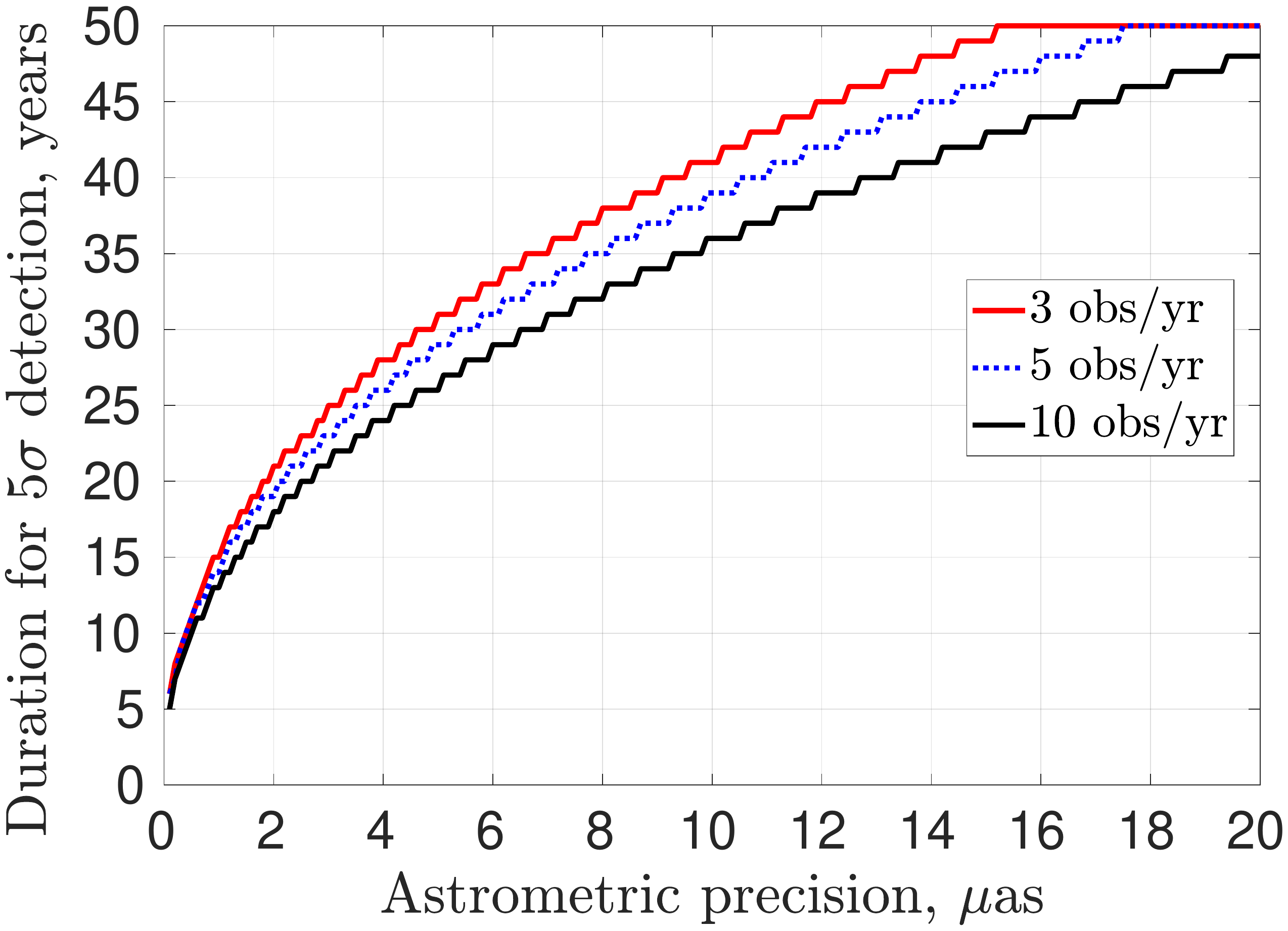}
	\includegraphics[width = 8.5cm] {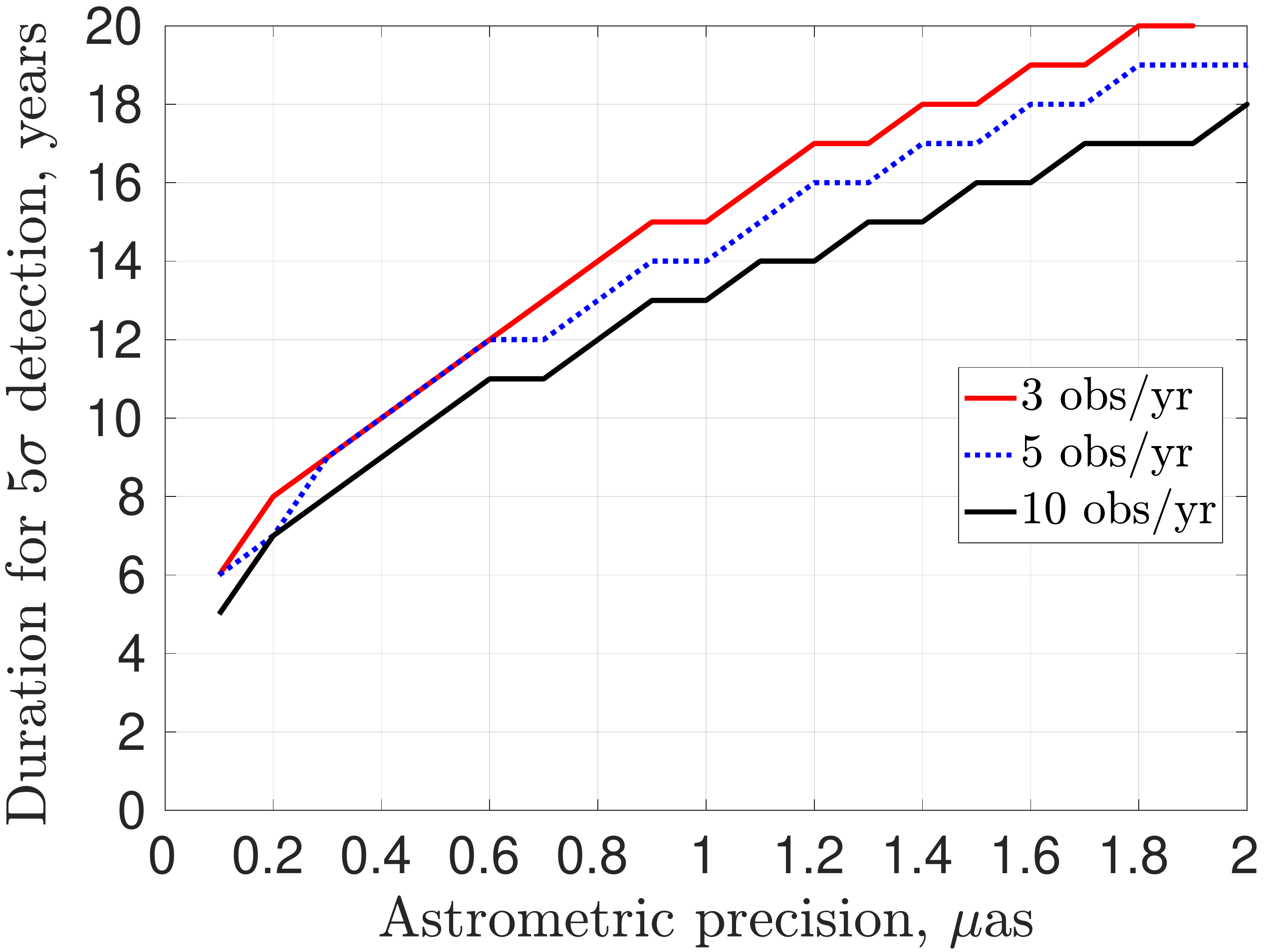}
	\caption{For a given individual epoch astrometric precision, the observing duration after which ${\geq 99\%}$ of Monte Carlo trials show a ${\geq 5}$ sigma discrepancy with Newtonian expectations. Results are shown for 3 successful annual pointings (red), 5 (blue dotted) and 10 (black). We only consider an integer number of years. A higher observing frequency modestly reduces the required duration. \emph{Left}: Results for precisions of ${\left( 0.1 - 20 \right) \, \mu}$as. The $y$-axis is cut off at 50 years, so a required duration of exactly this much implies that the proposed test is not feasible within 50 years at the indicated astrometric precision. \emph{Right}: Zoom into the ${\left( 0.1 - 2 \right) \, \mu}$as range of precisions.}
	\label{Required_duration_5_sigma}
\end{figure*}

We use Figure \ref{P_detection_3} to map out the probability $P_{detection}$ as a function of $\sigma$ and $T$, thereby revealing the trade-off between them. Our results allow us to determine the amount of time $T$ required to reach some threshold value of $P_{detection}$. The particular threshold adopted does not affect $T$ very much because, at fixed $\sigma$, there is a sharp change in $P_{detection}$ after a certain amount of time. In our work, we choose a threshold of 99\% so that null detection of the MOND signature would render the theory very unlikely.

Our corresponding estimates for $T$ are shown in Figure \ref{Required_duration_5_sigma}. The jagged nature of the curves arise because we only consider an integer number of years. For the case of 3 observations per year over 10 years, the required astrometric precision agrees rather well with our analytic estimate in Section \ref{Analytic_estimate}.

Using the proposed Theia mission \citep{Theia_2017}, the expected astrometric precision is $\approx 0.5 \, \mu$as given that P Cen has a V-band magnitude of 11 \citep{Benedict_1993}. If Milgromian dynamics is correct, our results show that observations would almost certainly be in 5 sigma tension with Newtonian dynamics after ${\approx 10}$ years of observations. Our estimates of the required duration $T$ follow the expected $\sigma^{2/5}$ scaling rather well.

\section{Systematics affecting the proposed test}
\label{Systematics}


\subsection{Perspective effects}
\label{Perspective_effect}

The relatively small distance to P Cen enlarges perspective effects, which were discussed e.g. in \citet[][section 3.2]{Shaya_2011} and \citet[][section 2.4]{Pittordis_2018}. This is because the same tangential velocity $v_t$ implies a changing angular velocity as the heliocentric distance $r$ of the system changes. Its angular motion also changes the way in which its velocity decomposes into a tangential component $v_t$ and a radial component $v_r \, -$ that part of the velocity which is radial at one moment is partly within the sky plane at another moment. Both effects imply an apparent astrometric acceleration\footnote{We mean here the physical acceleration that would produce an apparent astrometric acceleration of the same magnitude.} of order $v_t v_r/r$ if $v_t$ and $v_r$ have similar magnitudes, which is the case here (see below).

$v_t$ is already known to within ${\approx 1}$ m/s thanks to the long time baseline provided by combining data from Hipparcos and Gaia DR2 \citep{Kervella_2018}. With an astrometric precision of 1 $\mu$as per year, $v_t$ would be known to an accuracy of 0.6 m/s. The corresponding uncertainty on the apparent astrometric acceleration would be $2.7 \times 10^{-3} \, a_{_0}$ using the observed $v_r = -22$ km/s \citep{Kervella_2017}.

The excellent astrometry required for the proposed test would lead to a very precisely known downrange distance $r$ based on trigonometric parallax. Given that ${r = 1.30}$ pc \citep{Gaia_2018}, an astrometric accuracy of 10 $\mu$as implies $r$ would be known to a precision of ${\approx 10^{-5}}$. Assuming $v_t = 24$ km/s \citep[][table 1]{Kervella_2016}, this level of uncertainty in $r$ contributes only ${10^{-3} \, a_{_0}}$ to the final result.

A more serious issue might be uncertainty in $v_r$, which is only known to a precision of ${\approx 30}$ m/s \citep[][section 2]{Kervella_2017}. This is much poorer than the $\approx 1$ m/s accuracy of the HARPS instrument they used \citep{Gonzalez_2018}, mainly because the absolute value of $v_r$ is required to correct for perspective effects. This entails estimating the gravitational redshift and convective blueshift of P Cen. For finding exoplanets, these corrections are irrelevant as they only affect the zero-point of $v_r$.

A 30 m/s uncertainty in $v_r$ leads to an extra uncertainty of $0.15 \, a_{_0}$, which would need to be reduced significantly to enable the proposed test. We estimate that $v_r$ would need to be known at the ${\approx 10}$ m/s level or better, which should be feasible by the time this test is attempted. One route to progress is provided by the fact that close binaries should statistically have the same radial velocity. As the number of such binaries is expected to improve greatly in the Gaia era \citep{Zwitter_2003}, spectroscopic observations of them can be combined with the anchor provided by the Sun in order to yield much improved estimates of stellar radial velocity zero points. Some improvements are also expected from more precise spectroscopic redshift measurements, for instance using the recently installed ESPRESSO instrument \citep{Gonzalez_2018}.

\subsection{Undetected companions}
\label{Undetected_companions}

A planet of mass $M_p$ at distance $r_p$ from P Cen would cause it to accelerate by $GM_p/{r_p}^2$. Assuming $M_p = M_\bigoplus$ and requiring an acceleration $\geq 0.2 \, a_{_0}$, we see that the planet could be up to 27 AU from P Cen. At that distance, its orbital period would be 400 years. This is long enough that the orbital motion of the planet would not be significant over the observing duration $T$. As a result, the effect of such a planet could be degenerate with the expected MOND signal.

Fortunately, this is only true if the planet is in a particular sky-projected direction relative to P Cen. As perceived on our sky, the planet would need to be along the line segment connecting P Cen with $\alpha$ Cen and have a particular value of $M_p/{r_p}^2$. Otherwise, the planet-induced acceleration would not have the same direction and magnitude as the expected MOND effect. In this case, it should be clear that observers are looking at the influence of an additional object rather than a departure from Newtonian dynamics. Current radial velocity data do in fact reveal an exoplanet in the habitable zone \citep{Anglada_Escude_2016}. Combined with astrometry, this has allowed astronomers to rule out part of the parameter space available to other exoplanets \citep[][figure 14]{Kervella_2018}.

It is also possible to try and directly detect an exoplanet in the relevant sky region. Although challenging for terrestrial planets, larger gas giants could perhaps be directly imaged. A Jupiter mass planet would induce an acceleration of $0.2 \, a_{_0}$ at a distance of 485 AU, which is $6.2 \arcmin$ on the sky. Thus, ruling out contamination of the MOND signal by a Jupiter-mass planet would entail searching for its emission along a narrow band of length $6.2 \arcmin$. This band would be shorter for a lower mass planet, partly compensating for the fact that deeper observations would be required.

An additional distant planet in the Solar System would have a similar effect to a planet orbiting P Cen, in this case by accelerating the observatory rather than P Cen. In principle, these possibilities could be distinguished by observing other stars e.g. $\alpha$ Cen. However, even the proposed Planet Nine would have a mass of only ${\approx 10 \, M_\bigoplus}$ at a heliocentric distance ${\ge 200}$ AU \citep{Batygin_2016}. Accurate tracking of the Cassini orbiter around Saturn \citep{Matson_1992} rules out planets with a larger ratio of mass to distance cubed \citep{Fienga_2016, Holman_2016}. A ${10 \, M_\bigoplus}$ planet 200 AU from the Sun accelerates the inner Solar system by only ${0.037 \, a_{_0}}$, much less than the expected MOND signal. Moreover, it is likely that stronger limits on Planet Nine will be available by the time Theia is flown.\footnote{Section 2.6 of \citet{Batygin_2019} states that the planet must be ${\ge 370}$ AU away if its mass is ${10 \, M_\bigoplus}$.} If the planet is actually detected, it could be accounted for in the analysis.

\subsection{Astrometric microlensing}
\label{Astrometric_microlensing}

The observed sky position of P Cen could be affected by gravitational lensing. For this to occur in a way that is not corrected for, there must be an undetected object closer to us than P Cen and close to the line of sight towards it. The lensing effect of a point-like mass was considered in more detail in section 2.3 of \citet{Banik_2015_MCE}. Following their derivation, we define the angular Einstein radius $\theta_E$ within which a source is significantly magnified by the lens,
\begin{eqnarray}
	\theta_E ~\equiv~ \sqrt{\frac{4GM\left(1 - x \right)}{c^2 x D_S}} \, ,
\end{eqnarray}
where $G$ is the gravitational constant, $c$ is the speed of light in vacuum, $M$ is the mass of the lens and its downrange distance is a fraction $x$ of the observer-source distance $D_S$. We assume the lens mass is ${0.01 \, M_\odot}$ because a heavier lens might be detected in direct imaging. Placing the lens halfway to the source ($x = 1/2$) gives $\theta_E = 7.91$ mas. This is much more than the signal induced by MOND, so astrometric microlensing could potentially be significant \citep{Belokurov_2002}.

A microlensing event has a characteristic time dependence very different to the constant orbital acceleration of P Cen. This time dependence arises from the relative lens-source transverse velocity $v_t$. Combining this with $\theta_E$ defines a typical lensing timescale
\begin{eqnarray}
	t_E ~&=&~ \frac{x D_S \theta_E}{v_t} \\
	    ~&=&~ \frac{\sqrt{4GMD_S x \left(1 - x \right)}}{c v_t} \, .
\end{eqnarray}
Using the same lens parameters as before implies that $t_E = 10.7$ hours or 0.0012 years for a typical transverse velocity of $v_t = 20$ km/s. If a microlensing event is in progress, this would cause the sky position of P Cen to vary by ${\approx 8}$ mas over a period of ${\approx 11}$ hours. Such a large deviation would be easily discernible, allowing observers to wait out the event and use the remaining astrometric epochs in their analysis.

The above considerations indicate that the results should not be affected by `strong' microlensing events in which the minimum lens-source angular separation ${\la 10}$ mas. However, the sky position of P Cen could be affected at the $\mu$as level even for much larger sky separations $\beta$. In the limit where $\beta \gg \theta_E$, the unlensed and apparent images of P Cen are separated by an angle of \citep[][equation 10]{Banik_2015_MCE}
\begin{eqnarray}
	\delta ~=~ \frac{{\theta_E}^2}{\beta} \, .
	\label{Distant_deflection_angle}
\end{eqnarray}
Note that we have neglected the highly demagnified image because this is very much fainter than the normal image, whose brightness is nearly unaffected by the lens. In this limit, the astrometric effect of microlensing is much more important than the magnification it causes, because the latter falls off as $\beta^{-4}$ \citep[][section 2.3]{Banik_2015_MCE}.

Equation \ref{Distant_deflection_angle} indicates that $\delta \ga 1 \, \mu$as only if $\beta \la 1 \arcmin$. Therefore, only lenses within ${\approx 1 \arcmin}$ of P Cen can noticeably affect our inference on its orbital acceleration, and this only if the angle $\delta$ changes significantly over the ${T \approx 10}$ year period covered by the observations. We now consider the time variation of $\delta$ in more detail.

The test of gravity we propose is not affected by a constant $\delta$, but this is expected to vary with time as the angle $\beta$ changes. Assuming the unseen lens moves directly away from P Cen on our sky, the induced proper motion of P Cen's image has a magnitude of
\begin{eqnarray}
	\left| \dot{\delta} \right| ~=~ \frac{{\theta_E}^2}{\beta^2} \dot{\beta} \, .
\end{eqnarray}
A constant $\dot{\delta}$ can be absorbed by adjusting the initial transverse velocity of P Cen. However, a non-zero $\ddot{\delta}$ could in principle be degenerate with the physical acceleration of P Cen. When determining $\ddot{\delta}$, we get a term involving $\ddot{\beta}$ and a term involving ${\dot{\beta}}^2$. The ratio of these terms is $\ddot{\beta}\beta/{\dot{\beta}}^2$. Assuming the lens is not itself a binary and has a heliocentric radial velocity similar in magnitude to its transverse velocity, our results in Section \ref{Perspective_effect} indicate that both $\ddot{\beta}$ and ${\dot{\beta}}^2$ are of order ${v_t}^2/{D_S}^2$. Hence, the term involving $\ddot{\beta}$ is smaller by a factor of $\beta$, which we have seen is much smaller than unity for lenses that cause $\delta \ga 1 \, \mu$as. Therefore, we can approximate that
\begin{eqnarray}
	\left| \ddot{\delta} \right| ~=~ \frac{2{\theta_E}^2}{\beta^3} {\dot{\beta}}^2 \, .
\end{eqnarray}
For observations covering a duration $T$, the resulting accelerated angular displacement of P Cen is
\begin{eqnarray}
	\frac{1}{2} \left| \ddot{\delta} \right| T^2 ~=~ \frac{4GM \left(1 - x \right) \left(v_t T \right)^2}{c^2 \left(x D_S \beta \right)^3} \, .
\end{eqnarray}

In principle, this could be very large for small enough $\beta$. However, our preceding discussion indicates that such a situation would lead to $\delta$ changing with time in a more complicated manner over a shorter period. To estimate whether such higher-order effects would be detectable, we consider the third order behaviour of $\delta$. Using our preceding argument that higher order time derivatives of $\beta$ can be neglected, we get that
\begin{eqnarray}
	\left| \dddot{\delta} \right| ~=~ \frac{6{\theta_E}^2}{\beta^4} {\dot{\beta}}^3 \, .
\end{eqnarray}
The resulting cubic angular displacement is
\begin{eqnarray}
	\frac{1}{6} \left| \dddot{\delta} \right| T^3 ~=~ \frac{4GM \left(1 - x \right) \left(v_t T \right)^3}{c^2 \left(x D_S \beta \right)^4} \, .
\end{eqnarray}
This could be detected by including a cubic time dependence in the astrometric fits. Assuming that such an analysis can detect a cubic astrometric deviation exceeding some threshold angle $\sigma$, we get that
\begin{eqnarray}
	\beta ~\ge~ \frac{\sqrt[4]{4GM \left(1 - x \right)} \left(v_t T\right)^\frac{3}{4}}{x D_S \sqrt{c} \sqrt[4]{\sigma}} \, .
	\label{beta_min}
\end{eqnarray}
Assuming $\sigma = 5 \, \mu$as, this gives $\beta \ge 43 \arcsec$ for an observing duration of ${T = 10}$ years. A lower limit on $\beta$ arises because $\delta \propto 1/\beta$, so $\delta$ changes significantly over the time needed for $\beta$ to change significantly. For higher order effects to not be discernible, this must take longer than the observing duration $T$, implying a minimum value for the timescale ${\beta}/\dot{\beta}$.

Equation \ref{beta_min} implies that the accelerated angular displacement is limited to
\begin{eqnarray}
	\frac{1}{2} \left| \ddot{\delta} \right| T^2 ~\le~ \sigma^\frac{3}{4}\sqrt[4]{\frac{4 GM \left(1 - x\right)}{c^2 v_t T} }\, .
	\label{Maximum_microlensing_effect}
\end{eqnarray}
For our adopted parameters, the upper limit is $3.3 \, \mu$as, a limit which is also valid if the sky-projected lens-source separation and velocity are orthogonal instead of parallel. With a longer observing span, this limit decreases further while the actual MOND signal grows as $T^2$. Thus, astrometric microlensing should not significantly affect the results if observers analyze their data with this in mind.

Our discussion so far assumes the existence of a low-mass lens closer to us than P Cen and within a few arcminutes of it on our sky. This is not very likely because the relevant volume $V$ is rather small, as we now demonstrate. Assuming that $\beta$ is no more than $\sqrt{2}$ of its minimum value (Equation \ref{beta_min}) so that there is an appreciable lensing effect, integrating over all possible lens positions between Earth and P Cen gives
\begin{eqnarray}
	V ~&=&~ \mathrm{\pi} D_S \int_0^1 \left( \beta x D_S \right)^2 dx \\
	~&=&~ \frac{2 \mathrm{\pi} D_S}{3} \sqrt{\frac{4GM \left(v_t T \right)^3}{c^2 \sigma}} \, .
	\label{Microlensing_volume}
\end{eqnarray}
For the assumed parameters, this gives a volume of only $7 \times 10^{-8}$ pc\textsuperscript{3}. Gravitational microlensing surveys for free-floating and long-period exoplanets indicate that these are not very common, with at most 0.25 Jupiter-mass planets per star \citep{Mroz_2017}. Given that our nearest main sequence star is 1.3 pc away, it is not very likely that during the course of the Theia mission, a free-floating exoplanet will pass through such a tiny volume. Moreover, the volume is even smaller for a lower mass exoplanet than the ${\approx 10}$ Jupiter masses we assumed.

Therefore, it is extremely unlikely that astrometric microlensing hampers the ability of future observers to test gravity using astrometry of P Cen. Even if this were to occur and the lens was not detected in direct imaging, it would generally cause the sky position of P Cen to behave in a more complicated fashion. This could be discovered by analyzing the astrometric data in more detail, in particular by allowing for higher time derivatives. Such an analysis could still miss the microlensing, but in this case it could only have a rather small effect on the critically important second time derivative of the astrometry (Equation \ref{Maximum_microlensing_effect}). Moreover, the effect would likely be in a different direction to the expected MOND signal. Thus, it is very unlikely that the light from P Cen is appreciably deflected by an object closer to us in such a way that the time variation of the deflection angle hampers our ability to test gravity.

\subsection{Reference stars}
\label{Reference_stars}

Ideally, observers would measure the relative acceleration between P Cen and $\alpha$ Cen. Unfortunately, this is not possible because the Theia field of view is planned to be only $0.5^\circ$ \citep[][section 4.5.4]{Theia_2017}, much smaller than the 2.18$^\circ$ angle between them. As a result, the sky position of P Cen must be measured relative to other reference stars.

Fortunately, there are many possible reference stars with a comparable brightness to P Cen and within the proposed Theia field of view. An important issue is whether these stars provide a sufficiently stable reference frame. Their positions and proper motions will be known very accurately thanks to the Gaia mission \citep{Perryman_2001}, but their sky positions also accelerate. This acceleration would be rather large if caused by a close binary companion, an issue we assume can be avoided by cross-correlating reference stars with each other. Only those stars which appear astrometrically stable would be used as reference stars. Binaries could also be removed with multi-epoch radial velocity measurements. Fortunately, this should still leave plenty of reference stars to work with because ${\approx 1/2}$ of stars are expected to be isolated, like the Sun \citep{Duchene_2013}.

While binary reference stars can perhaps be avoided, even an isolated star experiences astrometric acceleration due to perspective effects \citep{Sozzetti_2005}. For a star with similar radial and tangential velocity of ${\approx 20}$ km/s and a distance of ${d = 200}$ pc, the apparent acceleration is $\approx v_t v_r/d = 0.54 \, a_{_0}$ (Section \ref{Perspective_effect}). However, the larger distance than P Cen means that this is actually a rather small effect. Expressed in angular terms, the deviation from uniform linear motion is $\approx \frac{1}{2} \frac{v_t v_r}{d^2} T^2 = 0.11 \, \mu$as over a period of $T = 10$ years. This is much smaller than the $7.18 \, \mu$as difference between the Newtonian and Milgromian astrometric accelerations of P Cen \citep[][section 9.1]{Banik_2018_Centauri}.

Since orbital and perspective accelerations both induce an astrometric deviation $\propto T^2$, their ratio cannot be improved further. However, perspective accelerations could be corrected for if $v_t$ and $v_r$ were known. Assuming conservatively that these will be known to a precision of 0.2 km/s or 1 percent, the reference stars would certainly be a reliable point of comparison for determining the sky position of P Cen.

Because a reference star would be much further away than P Cen, its inferred acceleration is resilient to a larger actual acceleration of the reference star. Even so, our results indicate that this could pose a problem if it exceeds ${\approx 10 \, a_{_0}}$. This threshold is larger for more distant reference stars, but even at a distance of only 200 pc, it already greatly exceeds the Galactic acceleration of a star in the Solar neighbourhood (Section \ref{Galactic_acceleration}).

Accelerations ${\ga 10 \, a_{_0}}$ might arise if the reference star was itself part of a WB. In this case, some of the acceleration would be along the line of sight. An acceleration of ${10 \, a_{_0}}$ entirely along this direction would cause the radial velocity to change by 38 cm/s over a decade, which might be discernible with accurate spectroscopic measurements. Moreover, the WB companion might be directly detectable.

Reference stars can also experience astrometric acceleration due to microlensing effects (Section \ref{Astrometric_microlensing}). The larger distance to the source means that there is a larger volume $V$ within which a lens would be problematic. However, Equation \ref{Microlensing_volume} indicates that $V \propto D_S$ such that it is very small even for a reference star many kpc away. Its maximum astrometric acceleration is in any case limited to much less than the expected MOND signal, regardless of how far away the reference star is (Equation \ref{Maximum_microlensing_effect}).

While astrometric stability of the reference stars is an important issue, we expect that this can be addressed with a modest amount of follow-up observations to determine the most suitable reference stars and measure their bulk motion. This can be used to correct for perspective effects, which are further reduced by using more distant reference stars. Doing so also reduces the astrometric effect of a given physical acceleration, making the results less sensitive to details of the reference star's environment. Thus, a combination of follow-up observations and careful selection of reference stars should reveal suitable ones that enable sub-$\mu$as astrometry of P Cen. Future Gaia data releases should provide much more information regarding possible reference stars.

\subsection{Finite size of Proxima Centauri}
\label{Finite_size_effects}

P Cen has an angular diameter of ${1.02 \pm 0.08}$ mas \citep{Segransan_2003, Demory_2009}.\footnote{Both works identify it using the alternative name GJ 551.} This may limit its astrometric precision because starspots can shift the photometric centre.  For Solar-type stars, several estimates exist for the average shift due to starspots \citep[e.g.][]{Sozzetti_2005, Eriksson_2007, Lagrange_2011}. The latter authors quantify an effect of order $0.1 \, \mu$as. Applying this to P Cen suggests that the effect of starspots is non-negligible but would not prevent the test of gravity proposed here. It is not currently known how significant jitter from starspots will ultimately prove to be for P Cen, but it is encouraging that its Gaia DR2 parallax is already accurate at the 0.2 mas level \citep{Gaia_2018}. The astrometric precision might be further improved by utilising concurrent photometric data as these will place constraints on possible starspots.

\subsection{Motion during each exposure}
\label{Blurring}

The astrometric position of P Cen is affected by its motion relative to the observatory. Assuming the observations are taken from a spacecraft on a similar orbit as the Earth, its heliocentric velocity would be ${\approx 30}$ km/s \citep{Hornsby_1771} while that of P Cen is 32.5 km/s \citep{Kervella_2016}. Depending on the exact geometry, the relative velocity would be ${\approx 50}$ km/s. If all of this were directed orthogonally to the line of sight, then the sky position of P Cen would change by $2.6 \, \mu$as over a 10 second exposure.

This blurring effect could be corrected for as it depends on the accurately known velocities of P Cen and the spacecraft. It can also be mitigated using shorter exposures, which are in any case a good way to avoid saturating the detectors given that P Cen is an 11\textsuperscript{th} magnitude star in the V-band \citep{Benedict_1993}. The combination of its high brightness and large across-scan motion will likely create unique data processing challenges \citep[][sections 3.3.5 $-$ 3.3.8]{Gaia_2016}.

\subsection{Galactic orbital acceleration}
\label{Galactic_acceleration}

The Galactic orbit of the whole $\alpha$ Cen system implies an acceleration of ${\approx 1.75 \, a_{_0}}$. This should be almost the same as the acceleration of the Sun along its Galactic orbit. There would be a small difference because $\alpha$ Cen is 1.3 pc from the Sun \citep{Kervella_2016}, but this is only a very small fraction of the Galactocentric distance of the Sun. Taking this to be 8.2 kpc \citep{McMillan_2017}, we expect the difference in Galactocentric accelerations to be $\approx \frac{1.3}{8200}\times 1.75 \, a_{_0} = 2.8 \times 10^{-4} a_{_0}$. Such a small effect is not relevant for our analysis.

In the non-linear MOND framework, the gravity of the Galaxy affects the internal dynamics of our nearest WB even in the complete absence of tidal effects. Our estimate of the predicted Milgromian acceleration of P Cen carefully takes this external field effect into account \citep[][section 2.1]{Banik_2018_Centauri}. The external field itself is known rather well based on the circular velocity at the Solar Circle and its distance from the Galactic Centre (see their section 3.6). As the resulting centripetal acceleration slightly exceeds $a_{_0}$, our results are sensitive to the detailed shape of the MOND interpolating function. This is now rather well constrained using a variety of extragalactic observations \citep[][section 7.1]{Banik_2018_Centauri}. In particular, the RAR is well fit using the `simple' interpolating function adopted here \citep{Famaey_Binney_2005}.

\section{Conclusions}
\label{Conclusions}

It is possible to directly test the behaviour of gravity at low accelerations using astrometry of the Sun's nearest star. The 13 kAU separation of P Cen from $\alpha$ Cen A and B implies an orbital acceleration below the MOND threshold \citep{Beech_2009, Beech_2011}. If MOND is correct, the acceleration would be 45\% larger than in Newtonian gravity. Over a decade, this would cause an astrometric anomaly of 7.18 $\mu$as and a radial velocity anomaly of 0.52 cm/s (Section \ref{Expected_acceleration}). While the latter is likely to remain undetectable, the former may be discernible with $\mu$as astrometry over several years. If 3 observations are conducted per year over ${T = 10}$ years at an individual epoch astrometric precision of $\sigma = 0.5 \, \mu$as, the observations would almost certainly (>99\% probability) be in 5 sigma disagreement with expectations based on Newtonian dynamics (Section \ref{Detectability}).

The required precision is very challenging but within the specifications of the proposed Theia mission \citep{Theia_2017}. Their figure 4.38 shows that a precision of 0.5 $\mu$as should be achievable for P Cen given its V-band apparent magnitude of 11 \citep{Benedict_1993}. In general, we expect that maintaining a similar level of distinguishability between Newtonian and Milgromian dynamics requires a precision $\sigma \propto T^{5/2}$.

In addition to precise astrometry, it will also be necessary to know the absolute radial velocity of P Cen within ${\approx 10}$ m/s to properly correct for perspective effects (Section \ref{Perspective_effect}). This is significantly better than the currently available precision of 32 m/s \citep{Kervella_2017}, but should be attainable by the time the proposed test is attempted. The 1.30 mas angular diameter of P Cen \citep{Segransan_2003, Demory_2009} might also limit the astrometric precision due to noise induced by starspots (Section \ref{Finite_size_effects}).

Another complication is the possible presence of a distant exoplanet with period long enough that its orbital acceleration changes only imperceptibly during the course of observations (Section \ref{Undetected_companions}). However, such a planet could only mimic a MOND effect if it has a particular ratio of mass to separation squared and a sky position very close to the line segment connecting P Cen with $\alpha$ Cen. Direct searches could also be undertaken for such a planet as the relevant sky area would be rather small and known in advance.

Achieving $\mu$as astrometric precision over several years will be a challenge, though one the Gaia mission indicates is far from impossible \citep{Perryman_2001}. Perhaps the most straightforward way is to fly a Theia-like successor \citep{Theia_2017}. If its design goals are achieved, ultra-precise astrometry of the Sun's nearest star could have far-reaching implications for fundamental physics and thus our understanding of the whole Universe.

\section*{Acknowledgements}

IB is supported by an Alexander von Humboldt postdoctoral fellowship. We thank Antonaldo Diaferio and Alessandro Sozzetti for helpful discussions on the observational aspects. We are grateful for timely comments by the referee Will Sutherland. The algorithms were set up using \textsc{matlab}$^\text{\textregistered}$.

\bibliographystyle{mnras}
\bibliography{THEIA_bbl}
\bsp
\label{lastpage}
\end{document}